# SPECIFICATION OF THE STATE'S LIFETIME IN THE DEVS FORMALISM BY FUZZY CONTROLLER

Dahmani Youcef[1] and Hamri Maamar[2]

[1]Department of Computer Science, University Ibn Khaldoun, Tiaret, Algeria
dahmani_y@yahoo.fr
[2]Laboratory LSIS, University Aix-Marseille, France
amine.hamri@lsis.org

## ABSTRACT

*This paper aims to develop a new approach to assess the duration of state in the DEVS formalism by fuzzy controller. The idea is to define a set of fuzzy rules obtained from observers or expert knowledge and to specify a fuzzy model which computes this duration, this latter is fed into the simulator to specify the new value in the model. In conventional model, each state is defined by a mean lifetime value whereas our method, calculates for each state the new lifetime according to inputs values. A wildfire case study is presented at the end of the paper. It is a challenging task due to its complex behavior, dynamical weather condition, and various variables involved. A global specification of the fuzzy controller and the forest fire model are presented in the DEVS formalism and comparison between conventional and fuzzy method is illustrated.*

## 1. INTRODUCTION

The modeling and simulation formalisms are used in order to understand, to represent, and specify the dynamic of complex systems [1]. Different methods and techniques have been created in order to improve their formulation. We distinguish two main categories: Analytic methods, and modeling and simulation methods [2]. Formally, a large variety of dynamic behaviors can be formulated mathematically. However the corresponding equations are unable to provide accurate results due to a lack of information for such systems and the complexity of their combination. To overcome this issue, modeling and simulation methods have been created. The modeling and simulation is based on an experimental frame [3,4], offering the possibility of predicting the behavior of complex systems. Various approaches were defined to treat the two phases of modeling and simulation, depending on either time-driven or event-driven systems.

Model and simulate discrete events deal with systems whose temporal and spatial behaviors are complex to be treated analytically. The DEVS formalism (Discrete EVent system Specification) is one of the common formalism used in the simulation of dynamical systems [5]. It is known for its modularity, expressivity [6], however, it based on constant piecewise input-output trajectories to simulate continuous dynamic systems [7,8]. In order to overcome this issue, many variants on DEVS were adopted by introducing appropriate theories such as the cellular automata [9], fuzzy logic etc.

The incomplete knowledge of certain systems involves vagueness and incompleteness. This point was studied by fuzzy logic [10,11,12]. The main difference with the conventional analytic methods is, firstly, it doesn't require a rigorous mathematical model to control a system. In the most cases, it uses knowledge of human operators to develop the controller, synthesizing the human operator actions. Secondly, its characteristic is the simplicity integration of subjective data in the controller. Its utilization is recommended when the drive system is imprecise.

This work aims to assess the states lifetime of a DEVS model by a fuzzy controller. A case study of forest fire propagation is done. Our example is based on this remark: "the duration of a wildfire spread at dry and windy time is necessarily shorter than that of a rainy and calm weather". Starting from this remark, we have tried to translate this observation by a fuzzy controller and simulated it via DEVS formalism.

The remainder of this paper is organized as follows. First, the section 2 briefly reviews a background on fuzzy logic and the DEVS formalism. Section 3 is devoted to the specification of the fuzzy controller in DEVS formalism. The fourth section illustrates our example of forest fire spread; we present its different variables and its formal description in DEVS. The fifth part presents results and at the end, in the sixth section, a conclusion, with future works, is given.

## 2. THE FUNDAMENTAL CONCEPTS

### 2.1. The Fuzzy Logic

#### 2.1.1. Linguistic Variable

The linguistic variable is a variable whose values are words or sentences in a natural or artificial language. It is characterized by quintuple (L,T(L),U,G,M). where L is the name of the variable, T(L) is the set of fuzzy sets (linguistic values), U is the universe of discourse, G is the syntactic rule and M its semantic [5,13]. The Figure 1 illustrates an example of the linguistic variable "velocity" with three terms: slow, middle and fast.

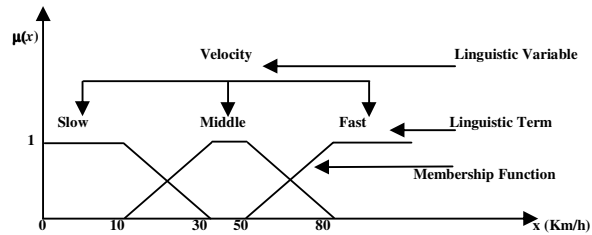

Figure 1. The membership functions of linguistic variable "Velocity"

#### 2.1.2. Linguistic Variable

The fuzzy rules [14] are expressions of this general form:

$$\mathbf{R_i} : \textbf{If } x_1 \text{ is } \mathbf{X_1^i} \text{ and } \ldots \text{ and } x_n \text{ is } \mathbf{X_n^i} \quad \textbf{Then } y \text{ is } \mathbf{Y} \tag{1}$$

Where $\mathbf{X_j^i}$ is a label of fuzzy set of the input $j$ ($j \in \{1..n\}$) and linguistic variable i ($i \in \{1..N\}$). Each linguistic term is characterized by its own membership function. Many forms can be used, trapezoidal, triangular, Gaussian (Figure 2).

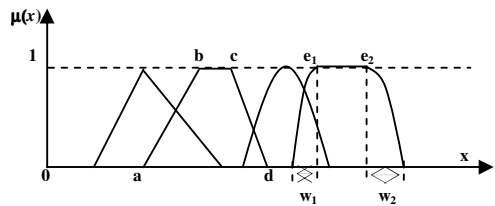

Figure 2. Common shapes of the membership functions

The consequent part is expressed by a fuzzy term, in this case, the rule is a Mamdani rule type, if the consequent is described as a function of its input variables, the rule is called Takagi-Sugeno rule type [11].

### 2.1.3. Fuzzy Inference System

A fuzzy inference system (FIS) also known as fuzzy controller aims to build a control law from linguistic and qualitative description of system's behavior via fuzzy rule base [15].
A Fuzzy controller is described by five main elements (Figure 3):
- Rule Base: Expresses the knowledge processes introduced by intuition and experimentation with Human operators.
- Data Base: Represents the properties of fuzzy sets.
- Fuzzification: Numerical values are transformed into linguistic variables with appropriate membership functions.
- Defuzzification: Transforms the command actions into crisp values useable directly by the controlled process.
- Inference Engine: Makes decisions through the activated fuzzy rules. It is the core of the controller.

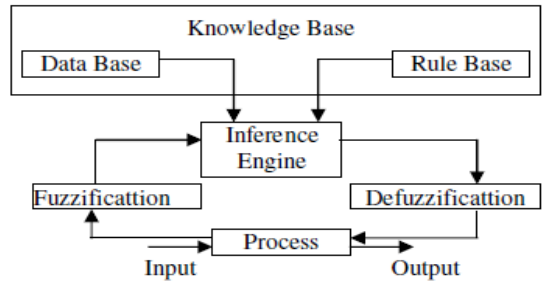

Figure 3. Fuzzy Inference System

### 2.1.4. Inference System

The inference is often reduced to the deduction in which the truth of the premises guarantees the completely truth of the conclusion. It is the decision-making mechanism; it gives the final conclusion for all activated rules according to the input data [14].
For an input vector $x=(x_1,\ldots,x_n)^t$, the fuzzy reasoning consists of 5 steps (Figure 4):
1. Obtain the membership degrees which match the appropriate membership function of each input.

$$\mu_{A_j^i}(x_j) \qquad (2)$$

2. Calculate the truth value of each rule.

$$\alpha_i(x) = \min_j \left( \mu_{A_j^i}(x_j) \right) \qquad (3)$$

3. Generate the contribution of each rule.

$$\mu(y) = \min \left( \alpha_i(x), \mu_{B^i}(y) \right) \qquad (4)$$

4. Aggregate the qualified rules.

$$\mu(y) = \max_i \left( \mu_{B^i}(y) \right) \qquad (5)$$

5. Produce the numerical value of the fuzzy output.

$$y = \frac{\int u\mu(u)du}{\int \mu(u)du} \qquad (6)$$

Where min stands for minimum function, max for maximum, n is the number of inputs whereas N is the number of fuzzy subsets.
This implementation is called "min, max, centre of gravity". It is the Mamdani inference method [10].

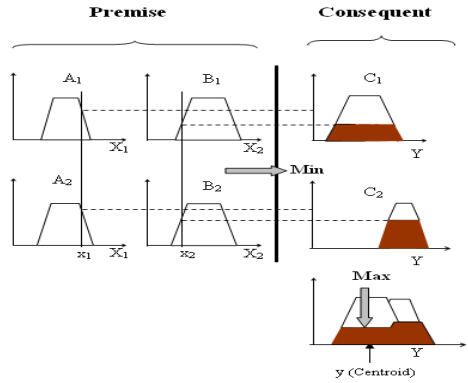

Figure 4. Fuzzy Inference

We recall that in the literature, many kinds of fuzzy reasoning exist. They depend on either the type of fuzzy rules or the nature of calculation the crisp output [14,16].

## 2.2. DEVS Formalism

### 2.2.1. Introduction

The DEVS formalism "Discrete EVent system Specification", was developed by Professor B.P. Zeigler [5]. It is based on mathematical theory of dynamic systems [7]. It is known for coupling heterogeneous models and separates the modeling process from the simulation one [8]. In fact this formalism is well adapted to represent a continuous system and describes the paradigm "event" in its overall features [13]. This formalism was applied in a great number of applications. It offers a general framework, and known as multi-formalism model [6].

Each system is characterized by two features: functional (behavioral) and structural aspects [9]. Similarly, the DEVS formalism authorizes two levels of description. At the lowest level, a basic part called atomic DEVS describes the behavior of a discrete event system. At the highest level, a coupled DEVS describes a system as modular and hierarchical structure [5,8].

### 2.2.2. The DEVS Atomic Model

The atomic models are the fundamental elements of the formalism; they describe the functional aspect of the system (Figure 5). They operate as "state-machines" [17]. Formally, a DEVS atomic model is described by seven-tuple (Equation 7):

$$AM = <X, S, Y, \delta_{int}, \delta_{ext}, \lambda, t_a> \qquad (7)$$

Where
X: the set of input events;

S: the set of partial states;
Y: the set of output events;
δ$_{int}$ : S→S : internal transition function, models the states changes caused when the elapsed time reaches to the lifetime of the state;
δ$_{ext}$ : Q×S→S : external transition function, defines how an input event changes a state of the system;
Q={(*s*,*e*) | *s*∈ S.0≤*e*≤t$_a$(*s*)} : total states and *e* describes the elapsed time since the last transition of the current state *s*;
λ: S→Y: when elapsed time reaches the state's lifetime, this function generates an output event;
t$_a$: S∈ R$_0^+$ ∪ ∞: time advance function, which is used to determine the lifespan of a state describing how long the system will stay in unchanged state if external events doesn't occur.

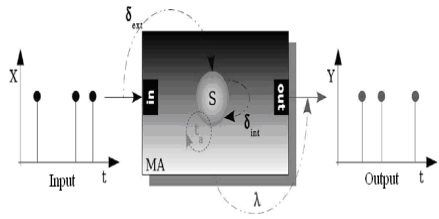

Figure 5. DEVS Atomic Model

### 2.2.3. The DEVS Couplded Model

The DEVS coupled model defines which sub-components belong to it and how they are connected to each other. It allows the creation of complex models starting from atomic and/or coupled models. Thus, it is modular and presents a hierarchical framework.
A DEVS coupled model is defined as an eight-tuple (Equation 8). A sample of coupled model is depicted on the Figure 6:

$$CM=<X_{self},Y_{self},D,\{M_d\},EIC,EOC,IC,Select> \qquad (8)$$

Where
Xself : set of possible inputs of the coupled model;
Yself : set of possible outputs of the coupled model;
D : is the name set of sub-components;
M$_d$ | d∈ D: set of sub-components which are either DEVS atomic or coupled model;
EIC: set of External Input Coupling;
EOC: set of External Output Coupling;
IC: defines the Internal Coupling;
Select: $2^D$→D: tie-break selector which select the event from the set of simultaneous events.

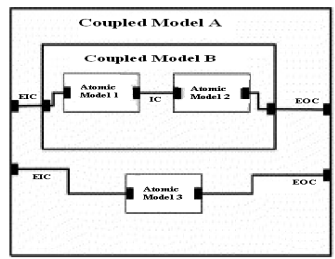

Figure 6. A simple DEVS Coupled Model

## 3. FUZZY-DEVS CONTROLLER

### 3.1. The Fuzzification Atomic Model

We assume that we have two variables $x_1$ and $x_2$ and a single output y. The linguistic terms of the variable $x_1$ are $A_1$ and $A_2$, while $x_2$, are $B_1$ and $B_2$ and those of y are $C_1$ and $C_2$. Therefore, the fuzzy rules are defined as follows:

**Rule $_i$**: If $x_1$ is $A_j$ and $x_2$ is $B_j$ Then y is $C_j$ with $j \in \{1,2\}$ and $i \in \{1..4\}$

As a rule base, we assume the table below (Table 1).

Table 1. Sample of Fuzzy Rule Base

| $x_1$ \ $X_2$ | $A_1$ | $A_2$ |
|---|---|---|
| $B_1$ | $C_1$ | $C_2$ |
| $B_2$ | $C_2$ | $C_1$ |

In the present work, every fuzzy set is depicted as trapezoidal shape (Figure 7).

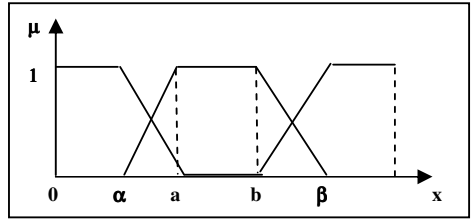

Figure 7. Trapezoidal membership function

Each membership function of the fuzzy inference system is considered as an atomic model. Its DEVS specification is defined by (Equation 9) and depicted in Figure 8:

$$\text{FuzzificationAM} = <X, S, Y, \delta_{int}, \delta_{ext}, \lambda, t_a> \quad (9)$$

Where
InPorts = {'InNum'}, $X_{InPorts} = \Re$
OutPorts = {'OutNum'}, $Y_{OutPorts} = [0, 1]$,
$X = \{(in, x) / in \in \text{InPorts}, x \in X_{InPorts}\}$,
$S = \{\text{'passive', 'active'}\} \times \Re$,
$Y = \{(out, y) / out \in \text{OutPorts}, y \in Y_{OutPorts}\}$,
$\delta_{int}(\text{'active'}, 0) = (\text{'passive'}, \infty)$,
$\delta_{ext}((\text{'passive'}, \infty), e, (\text{'InNum'}?x)) = (\text{'active'}, \mu(x))$,
$\lambda(\text{'active'}, m) = \text{OutNum}!m$
$t_a(\text{phase}, m) = 0 \quad$ if phase=active
$\qquad\qquad\qquad \infty \quad$ if phase = passive

$\mu(x)$ is the membership function (Equation 2) associated to the below model (Figure 8). The initial state of this model is (passive,∞).
For each input value, FuzzificationAM performs a calculation. The result represents the degree of membership to the associated fuzzy set. FuzzificationAM is independent from fuzzy inference, but it depends from the typical shapes of the membership functions (Figure 2).

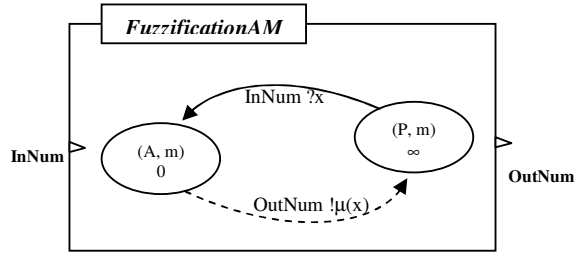

Figure 8. Fuzzification DEVS Atomic Model

### 3.2. Fuzzy Rule DEVS Atomic Model

According to the assumptions of section 3.1, each fuzzy rule has two inputs variables. Thus, the fuzzy rule is described as an atomic model (RuleAM) and its specification is illustrated as below (Equation 10):

$$\text{RuleAM} = <X, S, Y, \delta_{int}, \delta_{ext}, \lambda, t_a> \tag{10}$$

Where
InPorts = {'InNum1', 'InNum2'}, $X_{InPorts} = [0, 1]$
OutPorts = {'OutFuz'}, $Y_{OutPorts} = [0, 1] \times \Re^4$,
$X = \{(in, x) / in \in \text{InPorts}, x \in X_{InPorts}\}$,
$S = \{\text{'passive', 'active'}\} \times [0, 1] \times \Re^4$,
$Y = \{(out, y) / out \in \text{OutPorts}, y \in Y_{OutPorts}\}$,
$\delta_{int}(\text{'active'}, (\alpha, a, b, \beta), 0) = (\text{'passive'}, (\alpha, a, b, \beta), \infty)$,
$\delta_{ext}((\text{'passive'}, (\alpha, a, b, \beta), \infty), e, (('InNum1' ? x1) \& ('InNum2' ? x2))) = (\text{'active'}, x, (\alpha', a', b', \beta'))$
$\lambda(\text{'active'}, m, (\alpha, a, b, \beta)) = \text{OutFuz!} (m, (\alpha, a, b, \beta))$
$t_a(\text{phase}, m, (\alpha, a, b, \beta)) = 0$ if phase = active
$\quad\quad\quad\quad\quad\quad\quad\quad\quad\quad = \infty$ if phase = passive

x=min(x1,x2) which is given by Equation 2., while $(\alpha', a', b', \beta')$, is calculated by Equation 4.
The initial state of this model is (passive, $\infty$, ($\alpha, a, b, \beta$)).
When RuleAM receives x1 and x2 from FuzzificationAM, it transitions to active state otherwise it remains in passive state. The transition to the active state is conditioned by the occurrence of both inputs. The RuleAM depends on the rule base. It produces the contribution of each rule (step 3 of fuzzy inference) based on the outputs value of the FuzzificationAM.

### 3.3. Defuzzification DEVS Atomic Model

A defuzzification atomic model (DefuzzificationAM) outputs y. This value corresponds to crisp value which will be used to control the system. It is formally defined as:

$$\text{DefuzzificationAM} = <X, S, Y, \delta_{int}, \delta_{ext}, \lambda, t_a> \tag{11}$$

InPorts = {'InFuz'}, $X_{InPorts} = [0, 1] \times \Re^4$, OutPorts = {'OutNum'}, $Y_{OutPorts} = \Re$,
$X = \{(in, x) / in \in \text{InPorts}, x \in X_{InPorts}\}$, $Y = \{(out, y) / out \in \text{OutPorts}, y \in Y_{OutPorts}\}$,
$S = \{\text{'passive', 'active'}\} \times \Re$,
$\delta_{int}(\text{'active'}, 0) = (\text{'passive'}, \infty)$,
$\delta_{ext}((\text{'passive'}, \infty), e, \text{'InFuz'} ? (x1 \& x2 \& x3 \& x4)) = (\text{'active'}, u)$
$\lambda(\text{'active'}, m) = \text{OutNum!} m$
$t_a(\text{phase}, m) = 0$ if phase = active
$\quad\quad\quad\quad\quad = \infty$ if phase = passive

*u* is obtained by Equation 6, corresponding to defuzzification method. The DeffuzificationAM generates a final conclusion of the fuzzy controller based on the activated rules of the rule base. It begins with the passive state (passive, ∞) until receives all RuleAM outputs (four contribution rules, see section 3.1) otherwise none output will be done and the model remains in passive state. The DefuzzificationAM depends on the type of fuzzy inference adopted [18]. In our case, the inference employed is the centre of gravity.

### 3.4. FIS DEVS Coupled Model

As mentioned in section 3.1, we have used Mamdani rules type. Thus the fuzzy inference system coupled model (FIS_MamdaniCM) consists of 4 FuzzificationAM, 4 RuleAM, 1 DefuzzificationAM, 2 inputs and a single output. It is formally depicted in Figure 9.

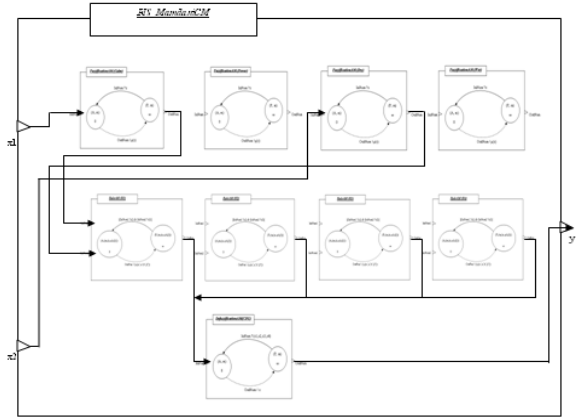

Figure 9.  FIS DEVS Coupled Model

## 4. FUZZY-DEVS CONTROLLER

### 4.1. Problem Identification

Due to the dynamic and complex nature of wildfire, it is impossible to identify, capture and model all influential parameters with absolute accuracy [19,20,21]. Thus, its formulation is very complex in terms of taking all its parameters. DEVS seems a useful tool and appropriate solution for this dynamic process. However in this formalism, each lifetime is a piecewise constant over the time, therefore any evolution in the environment will not appear on our modeled system. In this work, we try to give a solution for this issue by introducing a fuzzy controller to assess modification when the input events occur on the system.

The literature distinguishes three classes of parameters which set the fire spread ratio: vegetation type (caloric content, density...); fuel properties (vegetation size) and environmental parameters (wind speed, humidity and slope...) [22]. The forest fire evolves mainly according to the direction of the wind, its velocity and the relative humidity.

In the present work, we use two parameters: wind velocity (V) and humidity (H). We have identified five possible states that a cell can take (Figure 10). Each cell represents a limited area of the forest [23]:

- Nonflammable area (N): It can be a road, a surface of water or just an empty surface.
- Unburned area (U): Passive state; it represents any fuel which is not consumed yet by fire.
- Burning area (B): represents a consuming fire.
- Ember area (E): A small, glowing piece of coal or wood, as in a dying fire.
- Ash area (A): It is afterburning state; it is the final combustion process state. At this stage, the non-volatile products and residue were formed when matter is burnt.

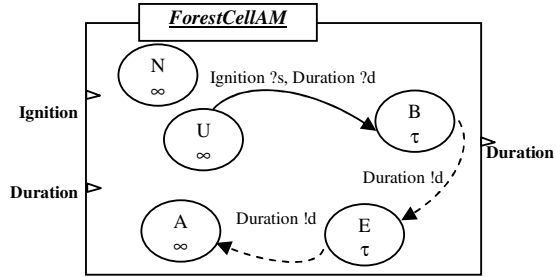

Figure 10. Forest cell DEVS atomic model

Each state's lifespan depends on the ignition and duration inputs values. The Ignition port indicates the fire start time (at what time the fire was triggered?), while the port Duration, it brings the consumption time of each forest cell.

### 4.2. Fuzzy Reasoning

According to our forest cell atomic model (Figure 10), we note H the relative humidity parameter, whereas V the wind velocity. The fuzzy logic controller describes the structure of the fuzzy rules as follows:

**Rule$_i$: If H is A and V is B Then $\tau_f$ is C**     (12)

A, B and C are linguistic variables and $\tau_f$ stands for fuzzy lifetime (fuzzy consumption time).
The variables are fuzzified as below (Figure 11).
The variable humidity H is divided into two fuzzy sets (linguistic term): Dry (D), and Wet (W). The wind velocity V is also fuzzified into two fuzzy sets: Calm (C), and power (P). The output variable $\tau_f$ is also fuzzified into two sets: Slow (S), and Fast (F).
The universe of discourse of each variable is given by:
- H: its values belong to [0%, 100%];
- V: is the interval [0,100km/h];
- $\tau_f$: The firefighters estimate the fire consumption of each cell at approximately 3 to 8% of the wind speed [24].

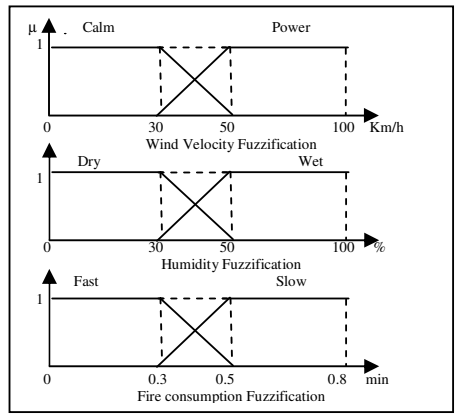

Figure 11. Fuzzification of variables H, V and $\tau_f$

The fuzzy rules base is given by Table 2. This table is filled by firefighters. It is obtained by their experiences.

Table 2. Experimental Fuzzy Rules Base

| $x_1$ \ $x_2$ | D | W |
|---|---|---|
| C | S | S |
| P | F | S |

The fuzzy inference system uses the method min-max centre of gravity. It calculates the consumption lifetime of each state and the result is provided to forest fire coupled model (Figure 12).

### 4.3. Fuzzy Reasoning

The proposed architecture is a classical DEVS framework. Our challenge is to keep the DEVS formalism unchanged and to improve it without modifying its components.

Our contribution is the addition of the FIS module whose function is to assess the lifetime of each state according to the input parameters: wind velocity (V) and humidity (H).

Initially, we fill the fuzzy rules base gotten from firemen reasoning. Each fuzzy rule is composed of two parts. The premise part, initially obtained from a data generator, and the consequent part which represents a state variable of the rule's DEVS atomic model.

The generator is a DEVS atomic model; it provides two kinds of values: spatial-temporal and environmental data. The spatial-temporal data are fed into forest coupled model, they supply the fire trigger event, while the environmental data, are fed into the FIS coupled model to compute the duration of fire consumption (Figure 12).

The forest coupled model is a grid composed of n lines and m columns. Each cell represents a forest cell atomic model (Figure 10) which is connected to its neighbors and provides the duration time obtained from the FIS coupled model. Each cell represents a DEVS atomic model which is associated to one simulator.

The dynamic system of the flaming front propagation speed is given by the simulator. It is based on the current cell position, consumption period and the wind direction. The wildland fire is considered as a propagation process where all burning cells ignite their unburned neighboring cells.

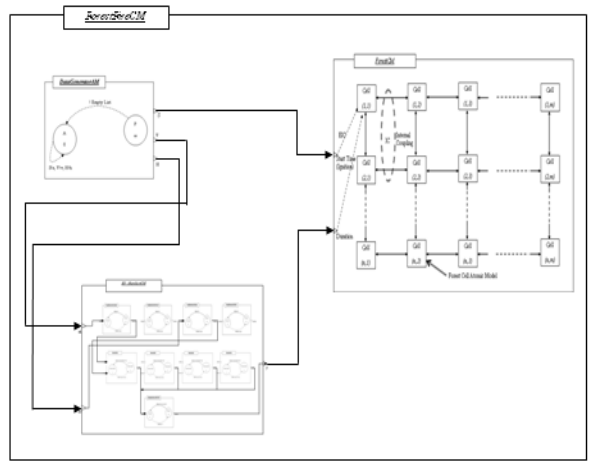

Figure 12. Forest fire DEVS coupled model

## 5. IMPLEMENTATION AND RESULTS

The simulator used in this work is implemented in Java. It is developed in LSIS laboratory. Much functionalities are inspired from its predecessor LSIS-DME [25]. This version lacks of visual modeling tool, however its utilization is very simple. The different paradigms of DEVS are defined as classes like root, simulator, atomic model, coupled model and so on. Each model inherits these classes and each implementation is easy to model despite the manner of construction.

### 5.1. Variables Setting

In order to test our approach, two kinds of simulations are done. In the first one, we assume the lifetime of each state as a piecewise constant value. In the second simulation, the lifetime is obtained by the fuzzy controller.
In these simulations, the different values are:
- Wind velocity: Its value is 35 km/h.
- Humidity coefficient: (45%).
- Wildland: Closely spaced.
- The fuzzy controller outputs the propagation velocity. For each cell, $\tau$ is obtained as an output of the atomic model described by Equation 11.
- Virtual forest is constructed as a grid of 90×90 cells where each cell represents an area of 1.2×0.8 m².
- Each cell is connected to 8 neighbors to form a coupled model. Nearest neighbors are defined as grid.
- The initial ignited cell is the cell (1,1) (Figure 12).
- We assume uniform parameters characterize the cell space, i.e. the direction and wind speed, and the humidity are constant along the forest fire area.

### 5.2. Results and Discussion

To compare the simulation performance between the conventional DEVS lifetime state and the fuzzy one, two experiments on forest fire propagation are executed using the parameters described in section 5.1. The difference concerns the manner to obtain the duration of each state.
The simulations were carried out on a Dell System GX280 with Intel ® Pentium (R) IV, CPU 2.80GHz processor,2G DDR2 SDRAM memory and Linux 2.6.32-5-686 operating system .
The Table 3 summarizes some important results. The model ForestFireSimZ uses a conventional lifetime while ForestFireSim uses our approach. In the latter model, an atomic model was added in order to compute the duration of the cells fire consumption. This addition ensures the obtaining of the duration depending on weather changes.

Table 3. Comparative Results.

| Results | Conventional DEVS lifetime | Fuzzy DEVS lifetime |
|---|---|---|
| Cell consumption time (Duration ($\tau$)) | 0.5 minutes | 0.556 minutes |
| Forest consumption time | 64.5 minutes | 69.6 minutes |
| Duration of the simulation | 616.29 seconds | 639.75 seconds |

To get better results, we have used additional free software which is Jconsole. It is a JMX-compliant monitoring tool. The table 4 resumes some important performances analysis between both models.

Table 4. Performance Results.

| Performance | ForestFireSimZ Model | ForestFireSim Model |
|---|---|---|
| Uptime | 10 minutes | 10 minutes |
| Process CPU time | 3 minutes | 4 minutes |
| Total compile time | 18.819 seconds | 3.688 seconds |
| Total threads started | 183,553 | 199,785 |
| Current classes loaded | 1,912 | 1,909 |
| Total classes loaded | 1,937 | 1,946 |
| Total classes unloaded | 25 | 37 |
| Current heap size | 14,345 kbytes | 9,083 |
| Committed memory | 17,380 kbytes | 18,428 |
| Total physical memory | 2,065,076 kbytes | 2,065,076 |
| Free physical memory | 616,392 kbytes | 607,060 |

According to these results, we remark that our approach brings some computation overhead compared to the traditional one. However, this method can add an interactive aspect by modifying the trajectory of the process without a great effort. It is sufficient to adapt the rules base and the lifespan of each state is modified immediately. However, a statistical study may be of interest to determine the compatibility of this comparison results and the viability of this approach.

## 6. CONCLUSION AND FUTURE WORK

For dynamic processes whose modeling accuracy requirements surpasses the classic discrete event specification that uses mean state lifetime, this work has presented an approach without modifying the core of the DEVS formalism and introduces the concept of interactive lifetime by showing the relationship between the input values and the duration of the states. This method allows adjusting the trajectory of the process even if the input values change. Also, it can ensure a dynamic structure of the model.

The structural and behavioural framework was developed and implemented. Some relevant results were presented at the end of this work.

We have applied this method on forest fire propagation. An overview was presented on the relevant parameters whose influence is considered important. We have adapted the DEVS formalism by allowing for uncertainties without modifying the structure of the classic DEVS specification.

Thereby, the resulting application simulates forest fire propagation, including imperfect data. A comparison between the traditional simulation and our approach was given. However, this work needs to be tested in real environment to judge its efficiency.

Many parameters remain to be introduced in this model as topology, inflammability etc. This addition will help in affirming the validity of our approach.

Our point of view is that the model presented here, to calculate the state lifetime by a fuzzy controller, can complement rather than compete with the more popular deterministic or stochastic DEVS models. In absence of a formal model, this process can be possible. Also the fuzzy lifetime function proposed in this paper is tentative, providing a satisfactory model for the forest fire is beyond the scope of this work.

**Authors**

**Youcef DAHMANI** is associate professor in department of computer science at University Ibn Khaldoun Tiaret. He obtained his diploma of computer engineering in 1992 from U.S.T.Oran, Algeria, and the MSc degree in 1997 from university of Es Senia Oran and received the PhD in 2006 from U.S.T.Oran. His research interests include Simulation Methodology, optimization of of fuzzy rules, Artificial Intelligence, reactive robotic systems and network security.

**Maamar HAMRI** is associate professor at Aix Marseille University and a member of LSIS laboratory. His main research is the discrete event simulation. Currently he is interested to the use of simulation in IA and software engineering. He is also member of the M&S network and supervises the M&S dictionary project.